\documentclass[aps,twocolumn,nofootinbib,preprintnumbers,eqsecnum,superscriptaddress]{revtex4-1}

\usepackage{color}
\usepackage{bbm}
\usepackage{tikz}
\usepackage{braket}
\usepackage{comment}
\usepackage{lipsum}
\usepackage{mathtools}
\usepackage{comment}
\usepackage{mdframed}
\usepackage[
pagebackref=false,
colorlinks=true,
linkcolor=blue,
urlcolor=blue,
filecolor=black,
citecolor=red,
pdfstartview=FitV,
pdftitle={},
pdfauthor={},
pdfsubject={},
pdfkeywords={},
pdfpagemode=None,
bookmarksopen=true
]{hyperref}

\usepackage[normalem]{ulem}
\usepackage{amsmath, amssymb, bm}
\usepackage{enumerate}
\usepackage{amsfonts}
\usepackage{epsfig}
\usepackage{mathbbol}
\usepackage{mathrsfs}
\definecolor{darkgreen}{rgb}{0.0, 0.5, 0.0}
\newcommand{\dg}{\dagger}
\newcommand{\ra}{\rightarrow}

\newcommand*{\AC}[1]{\textcolor{red}{[AC: \textsf{#1}]}}
\newenvironment{claim}[1]{\par\noindent\underline{\textbf{Claim:}}\space#1}{}
\newenvironment{conj}[1]{\par\noindent\underline{\textbf{Conjecture:}}\space#1}{}
\newenvironment{define}[1]{\par\noindent\underline{\textbf{Definition:}}\space#1}{}
\newenvironment{framework}[1]{\par\noindent\underline{\textbf{Framework:}}\space#1}{}

\begin{document}
	
	
	\title{The boundaries of 2+1D abelian fermionic topological orders}
	
	\author{Chang-Han Chen}%
	\affiliation{Department of Physics, Massachusetts Institute of Technology,
		Cambridge, MA 02139, USA}
	\email{cchan725@mit.edu}
	%
	\author{Xiao-Gang Wen}
	\affiliation{Department of Physics, Massachusetts Institute of Technology,
		Cambridge, MA 02139, USA}
	\email{xgwen@mit.edu}
	%

	
	\date{\today}
	
	\begin{abstract}

$2+1$D bosonic topological orders can be characterized by the $S,T$ matrices
that encode the statistics of topological excitations.  In particular, the
$S,T$ matrices can be used to  systematically obtain the gapped boundaries of bosonic
topological orders.  Such an approach, however, does
not naively apply to fermionic topological orders (FTOs).  In this work, we propose a
systematic approach to obtain the gapped boundaries of $2+1$D abelian FTOs.  The main trick is to construct a bosonic extension in which the fermionic excitation is ``condensed" to form the associated FTOs. Here we choose the parent bosonic topological order to be the
$\mathbb{Z}_2$ topological order, which indeed has a fermionic excitation.  Such a
construction allows us to find an explicit correspondence between abelian FTOs (described by odd $K$-matrix $K_F$) and the ``fermion-''
condensed $\mathbb{Z}_2$ topological orders (described by even $K$-matrix $K_B$).  This provides a systematic algorithm to obtain the
modular covariant boundary partition functions as well as the boundary
topological excitations of abelian FTOs. 
For example, the $\nu=1-\frac{1}{m}$ Laughlin's states have exactly one type
of gapped boundary when $m$ is a square, whose boundary excitations form a
$\mathbb{Z}_{2}\times\mathbb{Z}_{\sqrt{m}}$ fusion ring.  Our approach can be
easily generalized to obtain gapped and gapless boundaries of non-abelian
fermionic topological orders.


	\end{abstract}
	
	\maketitle
	
	\tableofcontents
	
	\section{Introduction }

	Condensed matter physics studies properties and organizations of all
kinds of materials. Perhaps not surprisingly, one of the central themes of
research has been the classification of phases of matters. Before the 80's, the
paradigm of \emph{symmetry breaking} proposed by Landau seemed to settle the
discussion: for example, all 230 kinds of crystals in three dimensions were
classified. 

	However, in the 80's, the discovery of the fractional quantum Hall
(FQH) effect clearly indicated the insufficiency of traditional
symmetry-breaking perspective.\cite{TSG8259,L8395} Even though the essence of
such phases are not fully understood yet, the notion of \emph{topological
orders} has been established,\cite{W8987,W9039,WN9077} via the ground state
degeneracy and the modular data (the $S,T$ matrices) which largely characterize
the 2+1D topological order.\cite{W9039,K062,RSW0777}

	The most straightforward approach is of course to write down the wave
functions. For example, the $\nu=\frac{1}{3}$ FQH states may be described by
\begin{align}
		\prod (z_i-z_j)^3 e^{-\frac{1}{4}\sum|z_i|^2},
	\end{align} 
the famous \emph{Laughlin's states}. In the past several decades, people have
successfully written down the wave functions of many FQH states, even the
non-abelian ones. Nevertheless, as the topological properties became more
understood, it also became clear that there may be a more algebraic approach to
describe the interplay between the topological excitations (so-called
\emph{anyons}). Along this line, the study of topological orders is a study of
the modular data of the anyons and the algebraic structure behind.
	
	One scheme that has been particularly useful is the theory of
topological order based on $(S,T,c)$.\cite{RSW0777,W150605768} Each of them
encodes different modular data of a $2+1$D bosonic topological order: $S$
matrix encodes the mutual statistics, $T$ matrix encodes the self-statistics,
and the chiral central charge $c$ encodes information about the gapless edge
modes. Of course, not every $(S,T,c)$ works; there must be some consistency
constraints, such as the Verlinde formula\cite{V8860} 
\begin{align}
		N^{ij}_k \equiv \sum_{l=1}^n \frac{S_{li}S_{lj}S_{lk}^\ast}{S_{l1}}\in \mathbb{N}.
	\end{align} 
It is an important, open question to find a complete set of constraints on
$(S,T,c)$. 

At a higher level, the theory based on $(S,T,c)$ is a special case of a more
general statement that topological orders in any dimensions are characterized
by their \emph{gravitational anomalies} on the boundaries. This is possible
because the bulk is a topological theory.  In two dimensions, the $S$ and $T$
matrices are representations of the mapping class group, under which the
boundary partition functions transform
covariantly\cite{W150605768,PhysRevResearch.1.033054}. Since this is a rather
modern perspective, let us sketch the logics in the following.

The mapping class group (MCG) is the automorphism group of the manifold,
quotioned out the "contractible" subgroup; in particular, MCG of the $n-$torus
is $\operatorname{MCG}(T^n)=\operatorname{SL}(n,\mathbb{Z}).$ It is widely
believed that anomaly-free conformal field theories (CFTs) are invariant under
MCG transformations (e.g., the modular invariance for 2D CFTs)\footnote{In
other words, the physics of a field theory should not depend on the
triangulation of the manifold, {\it i.e.,} the lattice realization; if it does,
then the theory is \emph{anomalous}.}, while CFTs with gravitational anomalies
are MCG covariant.   In other words, the states (and, consequently, the
partition functions) transform under a projective representation of MCG; each
partition function corresponds to the trace over a sector of the Hilbert space. 

More concretely, consider a CFT in a $d-$dimensional closed spacetime manifold
$M^d$ with gravitational anomalies. The partition functions $Z(g_{\mu\nu},i)$,
labeled by not only the metric $g_{\mu\nu}$ but also the index of sectors $i$,
transform as,
\begin{align}\label{ztrans}
	Z(g.g_{\mu\nu},i)=(R(g))_{i.j}Z(g_{\mu\nu},j),
\end{align} where $g\in \operatorname{MCG}(M^d)$ and $R$ is a projective representation of $\operatorname{MCG}(M^d)$. Since they transform covariantly, {\it i.e.,} like components of a vector, we call such object a \emph{multi-component} partition function.

As a side note, this is reminiscent of the t' Hooft anomaly of a global symmetry: given a global symmetry $G$, even though the operators transform linearly, {\it i.e.,}
\begin{align}
	U_gU_h\mathcal{O}(U_gU_h)^{-1} = U_{gh}\mathcal{O}U_{gh}^{-1},
\end{align} the states may transform projectively, {\it i.e.,}
\begin{align}
	U_{g}U_h\ket{\psi}=c(g,h)U_{gh}\ket{\psi}.
\end{align} The theory has a t' Hooft anomaly of $G$ if $c(g,h)$ cannot be removed by any UV regularizations.

Many t' Hooft anomalies can be realized on the boundary of symmetry-protected topological (SPT) orders.\footnote{It is not clear whether the edge/bulk correspondence holds for t' Hooft anomalies that are, {\it e.g,}  not phase-valued.\cite{Harlow}} Similarly, the gravitational anomalies may be realized on and classified by the topological orders in one higher dimensions. The edge/bulk correspondence is made explicit via the \emph{topological path integral}: the path integral can be performed over the bulk manifold with different anyon insertions, which, quotioned out the volume term, is a topological invariant. 

From the bulk perspective, the path integrals give rise to different topological ground states; from the boundary perspective, they are partition functions of different sectors. In this sense, the index $i$ in Eq.\ref{ztrans} also labels the anyon in the bulk, and thereby  
\begin{mdframed}
	\vspace{5 pt}
	the vector space of multi-component partition functions can be identified with the degenerate ground-state subspace of the bulk topological order.
	\vspace{5 pt}
\end{mdframed}

One can start from either side and gain knowledge on the other side. In this work, we focus on the $d=2$ case, and both the boundary and bulk theories are characterized by $(S,T,c)$.
	
	However, the theory of $(S,T,c)$ does not apply to fermionic
topological orders naively. The issue is whether we should identify the
electron with the trivial excitation. If we do, then the self-statistics would
have a $\pi$-ambiguity, {\it i.e.}, the $T$ matrix is not well-defined; if we
don't, {\it i.e.} treating the electron as a non-trivial excitation, then the
number of anyons would be doubled, and, consequently, the $S$ matrix becomes
non-invertible. More generally, the representation $R$ in Eq.\ref{ztrans} would not be unitary.

In the literature, this issue has been treated within the framework of
\emph{modular tensor categories} and its minimal modular extension, which is a
mathematically-intense
framework.\cite{GWW1017,LW150704673,BK160501640,LH200710562} In particular,
Ref. \cite{LH200710562} studied the boundary of fermionic topological orders.
Even though such algebraic approach may enable the classification of phases, it
somehow obscures the physics behind, especially the dynamics of anyons: for
instance, how to describe the real-time process of anyon condensation on gapped
boundaries? 

In this work, we adapt a simple ``hybrid'' approach. We resolve the
above issue by resorting to a bosonic extension of the fermionic system, based
on the philosophy of effective field theories. Guided by the physics intuition,
we propose the following framework: 
\begin{mdframed}
	\vspace{5 pt}
		\begin{framework}
			\begin{enumerate}
				\item Substitute the real electron with the emergent fermionic excitation of a simple bosonic topological order.
				\item To study the low-energy physics, restrict the Hilbert space to the ``local sectors'' of the bosonic topological order, that is generated by the emergent fermions only.
			\end{enumerate} 
		\end{framework}
	\vspace{5 pt}
	\end{mdframed}
Once we are in the bosonic-extended system, we carry our the $(S,T,c)$
treatment\cite{LWW1414,LW191108470}. In particular, we will compute the multi-component partition
functions in various examples.

	\section{Outline}

	In Sec.\ref{review}, we will briefly review some basic facts about the
Chern-Simons theory of abelian topological
order.\cite{W8951,FK8967,BW9033,FK9169,WZ9290} The key point is that the
Chern-Simons formalism efficiently encodes the modular data of abelian
topological orders in terms of the $K$ matrix. 

	In Sec.\ref{constuction}, we will demonstrate the embedding of a
fermionic system into the $\mathbb{Z}_2$ topological order via the $K$ matrix, namely Eq.\ref{kb}, inspired by the \emph{hierarchical construction}. 
We will show that there exists a canonical form of the "electron" as an emergent excitation in the new bosonic system (Eq.\ref{lele}) and how to
convert the excitations of the new system into the original ones (Eq.\ref{mapping}).  

	In Sec.\ref{bdry}, we will see that the $K$ matrix provides a straightforward algorithm to compute the $S,T$ matrices and thus the modular covariant partition functions of gapped boundaries, {\it i.e.}, lists of condensed anyons on the boundaries. We propose the following conjecture, 
		\begin{mdframed}
		\vspace{5 pt}
		\begin{conj}\\
			$Null(S-1)\cap Null(T-1)$ always has a rational basis.
		\end{conj}
		\vspace{5 pt}
	\end{mdframed}
We argue that a gapped boundary of the original fermionic system is given by the same list of condensed anyons, except restricted to the ``local sectors'', {\it i.e.}, having trivial mutual statistics with the electron. 
	
	In Sec.\ref{example}, we will apply the construction to several
variantions of the FQH states, and the results are consistent with the
Levin's\cite{L13017355}. Nevertheless, here we are able to compute the modular
covariant partition functions and consequently identify which anyons are
condensed. Three main results, based on the pattern in numerical data, are
\begin{itemize}
		\item The $\nu=1-\frac{1}{m}$ abelian FQH states \footnote{When $m$ is odd, this is the Laughlin's states formed by holes in the first Landau level. Our result applies to the $m$-even cases as well.} have exactly one type of gapped boundary\footnote{To be more precise, the vector space of multi-component partition functions is $1-$dimensional. But from now on, we will just use the phrase ``number of gapped boundaries" for convenience.} when $m$ is a square. 
The boundary topological excitations form a $\mathbb{Z}_{2}\times\mathbb{Z}_{\sqrt{m}}$ fusion ring.
		\item The double-$\nu=\frac{1}{m}$-Laughlin's states always have gapped boundaries. In particular, the number of gapped boundaries equals to $2$ when $m$ is a prime, and to $3$ when $m$ is a square. The boundary excitations form a  $\mathbb{Z}_{2}\times\mathbb{Z}_{m}$ fusion ring for all types of boundaries.
		\item More generally, the stacking of $\nu=\frac{1}{m}$ and $\nu=-\frac{1}{n}$\footnote{We will pick one of the Laughlin's states to be formed by the holes. Therefore, one may prefer to call such state a $\nu=-\frac{1}{n}$ FQH state. In this work, this is always implicitly assumed.} Laughlin's states have gapped boundaries when $mn$ is a square. We suspect that the boundary excitations form a $\mathbb{Z}_2\times\mathbb{Z}_{\sqrt{mn}}$ fusion ring for all types of boundaries.
	\end{itemize} We are less confident about the third point, for the reason explained in the paragraph after Eq.\ref{mark}. See Table.\ref{tab:stack} for the boundary fusion rings that we have computed.
	
	In Sec.\ref{alter}, we will give an alternative perspective, which hopefully justifies our framework. The key idea is to realize the vacuum as having $\mathbb{Z}_2$ topological order and identify a {\it trivial boundary condition}. This alternative approach will end up giving the same results for simple examples. For more complicated examples, this approach will be much less efficient, but we expect the results to be still consistent.
	
	Finally, to reiterate our framework, we believe that the boundary of
2+1D fermionic topological orders, including the non-abelian ones, may be
systematically studied by: \begin{itemize}
			\item finding a bosonic extension and its relations
to the original  fermionic topological order,
			\item and then identifying the electron and thus the low-energy/local part of the excitations for the  bosonic topological order
that reproduces all the
 excitations in the  fermionic topological order.
		\end{itemize}
	
	\section{Brief review on the Chern-Simons theory}\label{review}

	We give a minimum review on the Chern-Simons
theory\cite{Xiao:803748} that will be useful for us later. For simplicity and
concreteness, in this work we will focus on abelian fermionic topological
orders and embed them into the $\mathbb{Z}_2$ topological order. In this case,
the effective field theory can be described by the $U(1)^N$ Chern-Simons
theory: 
\begin{align}
		\mathcal{L} = -\frac{1}{4\pi}K_{IJ}a_{I\mu}\partial_\nu a_{J\lambda}\epsilon^{\mu\nu\lambda}+(...),
	\end{align} where $K$ is an invertible, symmetric $N\times N$ matrix and $(...)$ includes irrelevant terms, {\it i.e.} higher derivatives of the gauge fields, such as the Maxwell's terms. Such terms can be included when the dynamical properties are of interests\footnote{Take $\nu=\frac{1}{m}$ Laughlin's states for example. One can consider $\mathcal{L}=-\frac{m}{4\pi}a_\mu\partial_\nu a_{\lambda}\epsilon^{\mu\nu\lambda}+\frac{1}{2g_1}E^2-\frac{1}{2g_2}B^2$ and compute, say, equations of motion.}. 
	
	Each excitation of such theory is characterized by their gauge charges, labeled by a vector $l_I$. This amounts to adding a minimally-coupled term in the Lagrangian:
	\begin{align}
		\mathcal{L}_l = l_I a_{I\mu} j^\mu. 
	\end{align} This way, the Chern-Simons Lagrangian fully determines the modular data of anyons, {\it i.e.}, the algebraic properties of a topologically order. The self statistics is
	\begin{align}
		\theta_{self} = \pi l^T K^{-1} l
	\end{align} and the mutual statistics between two excitations, $l_1$ and $l_2$, is
	\begin{align}
		\theta_{1,2} = 2\pi l_1^T K^{-1} l_2.
	\end{align}
	
	Usually we also require the global $U(1)$ symmetry, in which case we
can assign a global $U(1)$ charge to the flux of each gauge field $a_{I,\mu}$,
labeled by the ``charge vector'' $q_I$, {\it i.e.} the $2\pi$ flux of  gauge
field $a_{I,\mu}$ carries an $U(1)$ charge $q_I$.  The simplest way to couple
such external $U(1)$ field, say, $A_\mu$ to the theory is then by including
\begin{align}
		\frac{e}{2\pi}q_I A_\mu \partial_\nu a_{I\lambda}\epsilon^{\mu\nu\lambda}
	\end{align} in the Lagrangian. Consequently, the charge of an excitation is given by
	\begin{align}
		Q_l = -eq^TK^{-1}l.
	\end{align} 
Later on, this formula will help us identify the ``electron'' with a fermionic
excitation of charge $-e$.  Furthermore, for FQH states, one can compute the
filling fraction 
\begin{align}\label{nu}
		\nu=q^T K^{-1}q
	\end{align}
	
	\section{Construction of the bosonic system}\label{constuction}
	Although there are, in principle, infinitely many $K$ matrices, we can relate a larger $K$ matrix to a smaller one by sequentially condensing excitations; this is known as the \emph{hierarchical construction}(see Appx.\ref{hiera} for a brief review). 
	
	Motivated by this construction, we extend a fermionic $K$ matrix to a bosonic one by treating the electron as the fermionic excitation of the $\mathbb{Z}_2$ topological order. The $\mathbb{Z}_2$ topological order is characterized by the $K$ matrix,
	\begin{align}
		K_{\mathbb{Z}_2}=\begin{pmatrix}
			0 &2\\
			2&0
		\end{pmatrix},
	\end{align} and has four excitations, usually called $\mathbb{1}, e, m$, and $f$. It is also conventional to identify the $e$ excitation with the \emph{gauge charge} of $\mathbb{Z}_2$ gauge theory, the $m$ excitation with the \emph{gauge flux}, and the $f$ excitation with the fermion, {\it i.e.,}
\begin{align}
	l_f =\begin{pmatrix}
		1\\
		1
	\end{pmatrix}.
\end{align} For more details, see Appx.\ref{hiera}
	\subsection{The form of $K_B$}  
	Consider a fermionic $K$ matrix,\begin{eqnarray}\label{kf}
		K_F = \begin{pmatrix}
			K_F{}_{1,1} & K_F{}_{1,2} &...\\
			K_F{}_{2,1} & K_F{}_{2,2} & ...   \\
			\vdotswithin{} & \vdotswithin{}& \vdotswithin{}&
		\end{pmatrix},
	\end{eqnarray} with the first element being odd, {\it i.e.,} $K_{1,1}\in 2\mathbb{Z}+1$, and all the others on the diagonal being even. Such $K$ matrix is a fermionic system. 
	
	We argue that a bosonic extension of $K_F$ is
	\begin{eqnarray}\label{kb}
		K_B &= \begin{pmatrix}
			0 & 2 & 1 &0  &...\\
			2 & 0 & 1 &0  &...\\
			1 & 1 & (1+K_F{}_{1,1}) & K_F{}_{1,2} &...\\
			0 & 0 & K_F{}_{2,1} & K_F{}_{2,2} & ...   \\
			\vdotswithin{} & \vdotswithin{}& \vdotswithin{}& \vdotswithin{}& \vdotswithin{}&
		\end{pmatrix}\\
	&=\begin{pmatrix}
		K_{\mathbb{Z}_2}&l_f&0&...\\
		l_f^T & (1+K_F{}_{1,1}) & K_F{}_{1,2} &...\\
		0 & K_F{}_{2,1} & K_F{}_{2,2} & ...   \\
		 \vdotswithin{}& \vdotswithin{}& \vdotswithin{}& \vdotswithin{}&
	\end{pmatrix}.
	\end{eqnarray}
	At the upper left corner is the $K$ matrix of the $\mathbb{Z}_2$ topological order and at the lower right corner is $K_F$ with the first element, {\it i.e.}, $K_{F1,1}$, incremented by $1$. The additional four $1$'s besides $K_{\mathbb{Z}_2}$ indicates the ``condensation'' of fermionic excitation (namely the $f$ excitation) that will be used for building the FQH state of $K_F$ via hierarchical construction. 
	
	When global $U(1)$ symmetry is imposed, we would consider the charge vector $q_F=(1,0,...,0)^T$ for the fermionic system, {\it i.e.} realized by single-layer FQH states, and let the corresponding bosonic one be $q_B =(-1,-1,0,...,0)^T$. One can generalize the situation to the multi-layer states. We would like to emphasize again that the global $U(1)$ symmetry is not required in our construction, but it may help us identify a new ``electron''.
	
	The ``$+1"$ in the $K_{B3,3}$ is the key of our construction. Now, the diagonal elements of $K_B$ are all even, which is a feature of bosonic topological orders. Therefore, we turn a fermionic system into a bosonic one, which gapped boundaries will be computed momentarily. Before that, we want to make sure that $K_B$ and $K_F$ are actually equivalent, when the appropriate equivalence relations are imposed. 
	
	It turns out that we should, as usual, require excitations to have
trivial mutual statistics with columns of $K_B$, which implies that each entry
of $l$ must be an integer (see Appx.\ref{proof1}). In addition, since the
quasi-particle wave-functions are single-valued, they must have trivial
statistics with the new ``electron''; otherwise, they are non-local 
with respect to the fermions and are considered as high-energy excitations, {\it
i.e.}, irrelevant to the lower-energy effective physics.
	
	\subsection{Equivalence relations of $l_B$}\label{eqrel} The
requirements above can be phrased more precisely as what follows:
	\begin{mdframed}
		\vspace{5 pt}
		\begin{define}
			\begin{enumerate}
				\item The ``electron'' is $l_{ele}=(1,1,1,0...,0)^T$, {\it i.e.} has the fermionic self-statistics (and charge $=-e$ when global $U(1)$ symmetry is imposed).
				\item 
The excitations are labeled by any integer vectors, $l_B\in \mathbb{Z}^n$, that have trivial mutual statistics with $l_{ele}$.
				\item The equivalent classes of $l_B$'s are defined by the following equivalence relations:
				\begin{itemize}
					\item $l_B\sim l_B+K_{B,i}$
					\item $l_B\sim l_B+l_{ele}$.
				\end{itemize}
			\end{enumerate}
		\end{define}
	\vspace{5 pt}
	\end{mdframed}
We will show that such definition of the excitations of the bosonic extension $K_B$ reproduce that of the fermionic topological order $K_F$.

	It is evident from Appx.\ref{proof} that the mutual statistics are invariant under such equivalence relations. This means that our definition of equivalence class of $K_B$ makes sense: topological excitations in the same class are indistinguishable by remote operations, hence ``physically'' equivalent\footnote{One can imagine probing the system through the braiding of anyons, in which the physical content is fully characterized by the statistics.}. 
	
	It is very important to notice the difference between our equivalence relations of $l_B$ and the usual ones of $l_F$. Usually, given a $K_F$, the equivalence relations are only given by 
	\begin{align*}
		l_F\sim l_F+K_{F,i},
	\end{align*} without mentioning anything about $l_{ele}$. This is because the electron is already a column of $K_F$\footnote{namely $K_{F,i}$, where $(K_{F})_{i,i}$ is odd.} and thus being treated as trivial. For $K_B$, we have to manually impose such equivalence relation via $l_{ele}$. 
	
	Since we claim that this bosonic system is dual to the fermionic one, there should be a \emph{natural identification} between the equivalence classes of $l_F$ and those of $l_B$. In practice, if such identification does not exist, then one needs to pick another $l_{ele}$ for the ``electron'' of the bosonic system. We will show that picking 
	\begin{align}\label{lele}
		l_{ele}=\begin{pmatrix}
			1\\
			1\\
			1\\
			0\\
			\vdotswithin{}\\
			0
		\end{pmatrix}
	\end{align} always works. Furthermore, given such $l_{ele}$, the identification becomes as simple as
	\begin{align}\label{identif}
		[l_B=(0,0,l_1,..,l_n)^T]\Longleftrightarrow [l_F=(l_1,...,l_n)^T],
	\end{align} where $[l]$ is the equivalence class that contains $l$.
	The identification should serve two purposes:
	\begin{enumerate}
		\item preserve the statistics, {\it i.e.}
		\begin{align}\label{stat}
			{l'_B}^T K_B^{-1}l_B = {l'_F}^TK_F^{-1}l_F,
		\end{align} if $l_B=(0,0,l_F)$ and $l_B'=(0,0,l_F')$.
		\item be consistent with both the equivalence relations of $l_B$ and of $l_F$.
	\end{enumerate}
	Eq.\ref{stat} is proved in Appx.\ref{statpres}, where the explicit form of $K_B^{-1}$ is provided. As a side note, this object may be of interest if one prefers the gauge-flux basis over the gauge-charge basis.
	
	To prove the second statement, we make the following claims and prove them in the same order:
	\begin{mdframed}
		\vspace{5 pt}
		\begin{claim}
			\begin{enumerate}
				\item To have mutual statistics with $l_{ele}$, any allowed $l_B=(l_1,l_2,...,l_n)^T$ has to satisfy 
				\begin{align}\label{even}
					l_1+l_2\in 2\mathbb{Z}.
				\end{align}
				\item For any such $l_B$, there exists 
				$l'_3\in\mathbb{Z}$ such that 
				\begin{align}\label{mapping}
					(l_1,...,l_n)\sim (0,0,l_3',l_4...,l_{n}).
				\end{align}
				We call $(l_3',l_4...,l_{n})$ the \emph{associated $l_F$ of $l_B$}.
				\item Two $l_B$'s are inequivalent if and only if the associated $l_F$'s are inequivalent.
			\end{enumerate}
		\end{claim}
	\vspace{5 pt}
	\end{mdframed}
	Eq.\ref{even} is proved in \ref{proof2}. To prove the second statement\footnote{The main point here is that we can make the first two entries zero via equivalence relations. One may notice that only $l_3$ need to be changed into $l_3'$ and the ones after $l_3$ don't. This is a cute coincidence but not relevant to our main arguments. }, If both $l_1$ and $l_2$ are even, then it is evident that we can use the first two columns of $K_B$, namely
	\begin{align*}
		\begin{pmatrix}
			0\\
			2\\
			1\\
			0\\
			\vdotswithin{}
		\end{pmatrix}\text{ and }
		\begin{pmatrix}
			2\\
			0\\
			1\\
			0\\
			\vdotswithin{}
		\end{pmatrix},
	\end{align*} to set $l_1=l_2=0$; if both are odd, we can first add an electron, so that $l_1\ra l_1+1$ and $l_2\ra l_2+1$, and then set both to zeros as in the previous case. 
	
	For the third claim, consider any integer linear combinations of  $l_{ele}$ and of columns of $K_B$. These are all the trivial excitations of $K_B$ (including the electron). As just argued, we can set the first two entries to be zero. Now, we want to show that they must be integer linear combinations of $(0,0,K_{F,i})^T$, where $K_{F,i}$ is the $i$-th column of $K_F$. It suffices to check this for the integer linear combinations of first three columns of $K_B$ and of $l_{ele}$, namely
	\begin{align*}
		\begin{pmatrix}
			0\\
			2\\
			1\\
			0\\
			\vdotswithin{}
		\end{pmatrix},
		\begin{pmatrix}
			2\\
			0\\
			1\\
			0\\
			\vdotswithin{}
		\end{pmatrix},
		\begin{pmatrix}
			1\\
			1\\
			1+K_{F1,1}\\
			K_{F2,1}\\
			\vdotswithin{}
		\end{pmatrix},
		\begin{pmatrix}
			1\\
			1\\
			1\\
			0\\
			\vdotswithin{}
		\end{pmatrix}.
	\end{align*} Some simple algebra shows that they are either $0$ or indeed $(0,0,K_{F,i})^T$. Therefore, if the two associated $l_F$'s are inequivalent, then two $l_B$'s must be inequivalent as well. The converse is obvious since the equivalence relations of $l_B$'s naturally contains those of $l_F$'s. This concludes the proof of the third claim. 
	
	\subsection{Some remarks} \label{remarks}
	Let us spell out the identification even more explicitly. First, to find the associated $l_F$ of a $l_B$, one first set the first two entries to zero and then use Eq.\ref{mapping}; conversely, to find a $l_B$ associated with a $l_F$, one can simply pick $l_B=(0 , 0, l_F^T)^T$. Of course there are some other less natural choices of $l_B$'s, but they are all equivalent as proved above. Anyways, for our current purpose, the important direction is to convert $l_B$ into $l_F$, and the ambiguity of converting $l_F$ into $l_B$ should not bother us.
	
	Second, having such natural identification implies that: 
	
	\vspace{.1 in}
	\begin{mdframed}
		\vspace{5 pt}
		\hspace{0.3 in}
		\begin{minipage}{\textwidth}
			The number of equivalence classes of $l_B$ \\
			is $|\det K_F|$ ($=\frac{1}{4}|\det K_B|$).
		\end{minipage}
	\vspace{5 pt}
	\end{mdframed}
	
	\noindent This is because given an $n\times n$ full-rank, integer matrix $K$, the number of distinct integer vectors up to addition of columns of $K$ \footnote{Notice that this is identical to the equivalence relations of $l_F$'s.} equals to $|\det K|$. One way to understand this is that $K$ defines a lattice, where the columns for the basis vectors. The volume of its fundamental parallelepiped is $|\det K|$, and each integer point inside the fundamental parallelepiped labels an allowed excitation. Since the numbers of equivalence classes of $l_B$ and of $l_F$ are the same, the statement follows.
	
	Finally, let's come back to the issue mentioned in the introduction: for a fermionic topological order, when the number of excitations equals $|\det K_F|,$ the $T$ matrix is ambiguous; when the number equals $2|\det K_F|$, the $S$ matrix is non-invertible. Now, in our $K_B$ construction, the number becomes $|\det K_B|=4|\det K_F|$, and the theory is a totally fine bosonic topological order with a fermionic excitation. $l_{ele}$.
	
	If we only consider the equivalence relations of a bosnoic topological order, {\it i.e.,} columns of $K_B$, then the $S$ and $T$ matrices are unitary and unambiguous. Only when we include $l_{ele}$ in the equivalence relations does the theory become dual to a fermionic topological order, and the same ambiguity issue arises. Therefore, throughout the calculation later, we will forget about $l_{ele}$ and compute the boundaries from bosonic $S$ and $T$ matrices. We invoke $l_{ele}$ in the end to obtain the fermionic boundaries from the bosonic results.
	
	\section{Boundary of the bosonic dual}\label{bdry}

	\subsection{Compute $S$ and $T$  from $K$}

	We have described how to obtain $K$ matrix, which efficiently encode
the modular data of anyons. In order to make the data more accessible, we would
like to compute the $S$ and $T$ matrices from the $K$ matrix. This is possible
because the wave function of the multi-layer FQH states characterized by a $K$
matrix can be written as\cite{Xiao:803748} 
\begin{align}\label{wf}
		\prod_{I;i<j}(z^I_i-z^I_j)^{K_{II}}\prod_{I<J;i,j} (z^I_i-z^J_j)^{K_{IJ}}e^{-\frac{1}{4}\sum_{i,I}|z^I_i|^2},
	\end{align} where $z_i^I =x_i^I+iy_i^I$ is the coordinate of the $i$th particle in the $I$th condensate. This confirms that the microscopic information is encoded in the $K$ matrix. 
	
	Second, recall that the elements of the $S$ matrix encode the mutual statistics, while the elements of $T$ matrix encode the self-statistics. From Eq.\ref{wf}, we can write $S$ and $T$ in terms of $K$:
	\begin{align}
		&S_{a,b} = \frac{e^{-i2\pi l_b^T K l_a}}{\sqrt{|\det K|}}\\
		&T_{a,b} = e^{-i2\pi\frac{c}{24}} e^{i\pi l_a^TKl_a}\delta_{a,b},
	\end{align} where $c$ is the chiral central charge and $a,b = 1,2,...,\det K$. The indices label the excitations, {\it i.e.}, $l_a$ is the gauge charges of the $a$th excitation. Of course, $T$ is diagonal in such basis.

For $\mathbb{Z}_2$ topological order, $K=K_{\mathbb{Z}_2}$, the $S$ and $T$ matrices are therefore
\begin{align}\label{stz2}
	S_{\mathbb{Z}_2}& = \frac{1}{2}\begin{pmatrix}
		 1& 1 & 1 & 1\\
		 1 & 1 & -1 & -1\\
		 1 & -1 & 1 &-1\\
		 1 & -1 & -1 & 1
	\end{pmatrix},\\
T_{\mathbb{Z}_2}& = \begin{pmatrix}
	1 & 0 & 0 & 0\\
	0 & 1 & 0 & 0\\
	0 & 0 & 1 & 0\\
	0 & 0 & 0 & -1
\end{pmatrix}.
\end{align} It should not be surprising that they are $4$-dimensional, since that is exactly the number of anyons of $\mathbb{Z}_2$ topological order. Indeed, the order of the basis is $\{\mathbb{1},e,m,f\}$
	
	\subsection{Modular covariant partition functions}
	Due to the gravitational anomalies, the partition functions form a vector, namely the multi-component partition function, transforming covariantly under the $S$ and $T$ matrices. As in 2D CFTs on a torus, the shape of $T^2$ is parametrized by a complex number $\tau$. Thus we write 
	\begin{align}
		Z(\tau,\overline{\tau};i)\equiv Z (\tau,\overline{\tau};\ket{\psi_i})
	\end{align} to denote the multi-component partition function. $\ket{\psi_i}$ is one of the degenerate ground states, associated with the type-$i$ excitation in the bulk.
	
	In two dimensions, Eq.\ref{ztrans} becomes
	\begin{align}
		S_{ij}Z(\tau,\overline{\tau};j) &= Z(-1/\tau,-1/\overline{\tau};i),\nonumber\\
		T_{ij}Z(\tau,\overline{\tau};j) &= Z(\tau+1,\overline{\tau}+1;i).
	\end{align}
Our goal here is to find the solutions for gapped boundaries, where each entry is just an integer independent of $\tau$.\footnote{If one hopes to describe the gapless boundaries, then they can try to construct such covariant functions out of the characters $\chi(\tau)$ and $\overline{\chi}(\overline{\tau})$ of rational CFTs.} In other words, the partition function is really like a constant, integer vector with dimension $|\det{K}|$, say $\textbf{Z}$, such that,
	\begin{align}
		&S\textbf{Z}=\textbf{Z}\nonumber\\
		&T\textbf{Z}=\textbf{Z}.\label{stv}
	\end{align}
  Therefore, the problem of finding partition functions of gapped boundaries amounts to finding positive, integer null vectors of $(S-1)$ and $(T-1)$.
  
	 If an entry is non-zero, then the associated excitation is condensed on the gapped boundaries. In the following we describe an algorithm to find such partition functions:
	\begin{enumerate}
		\item Find a basis for the null space of $S+S^\dg+T+T^\dg-4$. Notice that this basis may not be rational.
		\item Suppose that the dimension of the null space is $d$. Construct an $d\times\det K$ matrix, $V$, such that each row is a distinct basis vector. In other words,
		\begin{align}
			V=\begin{pmatrix}
				v_1^T\\
				\vdotswithin{}\\
				v_d^T
			\end{pmatrix}, 
		\end{align} where $\{v_i\}$ is a basis of the null space. 
		\item Use the Gaussian elimination\footnote{Here we say ``Gaussian elimination'' to illustrate the idea. In practice, our program uses the Smith normal form to increase the efficiency.} to obtain a new matrix $V'$, where there is a $d\times d$ identity block, {\it i.e.},
		\begin{align}
			V'=\begin{pmatrix}
				\cdots & 1 & 0  &\cdots&0&\cdots\\
				\cdots & 0 & 1  &\cdots&0&\cdots\\
				& \vdotswithin{}& &\ddots& &\\
				\cdots & 0 & 0  &\cdots&1&\cdots
			\end{pmatrix}
		\end{align}Notice that this only involves row operations.
		\item The claim is that now, each row of $V'$ is rational. 
We then scale each row of $V'$ to make it integral.
		\item Check whether these integral null vectors satisfy Eq.\ref{stv}, {\it i.e.,} are null vectors of $(S-1)$ and $(T-1)$ instead of just $S+S^\dg+T+T^\dg-4$.
		\item Find linear combinations of the integral null vectors,
such that they are non-negative, integral, and ``indecomposable" (elaborated
below).\footnote{For now, our program does this step by brute force. It is
desirable to have a better algorithm.} Notice that the coefficients of the
linear combination can be negative or fractional, as long as the outcome is
non-negative and integral. These are the solutions of interest.

	\end{enumerate}

	To prove the claim in step $4$, we need to assume the following conjecture:
	\begin{mdframed}
		\vspace{3 pt}
		\begin{conj}
			$Null(S-1)\cap Null(T-1)$ always has a rational basis.
		\end{conj}
	\vspace{3 pt}
	\end{mdframed}
	Given this conjecture, the proof goes as what follows. Each row of $V'$ is a linear combination of $d$ rational vectors. Notice that each row of $V'$ contains at least $d$ rational numbers, namely the $0$'s and $1$'s in the identity block. This is a system of $d$ linear equations, which fixes the coefficients of linear combination to be rational. The same argument works for each row of $V'$. Hence, $V'$ is rational.
	
	In the final step, we want to find all the non-negative and
``indecomposable vectors" in the null space. Here an non-negative
indecomposable vector is the one that cannot be written as a non-negative
linear combination of any other non-negative indecomposable vectors in the null
space.  Physically, the indecomposability means that a boundary cannot be
trivially realized as the stacking of the other two, so they correspond to
stable boundaries.  Since the coefficients of the linear combination can be
negative or fractional, the number of indecomposable boundaries can {\it a
priori} be larger than the dimension of the null space. Nevertheless, in the
simple cases of abelian topological orders that we tried, these two number
coincide. Whether or not these two numbers are identical mathematically, we
emphasize that only the notion of non-negative indecomposable boundaries are
physically meaningful, but not an arbitrary basis of the null space. 
	
	From now on, we will adopt a slight abuse of notations. When we say
``the number of gapped boundaries," we actually mean the number of
non-nagentive indecomposable vectors in the null space, which happens to coincide with the
dimension of the null space in all examples. 


	As an important example, let's compute the gapped boundaries of
$\mathbb{Z}_2$ topological order. The dimension of the null space is $2$.
Indeed, one can check the following two partition functions, 
\begin{align}
		\mathbf{Z}=\begin{pmatrix}
			Z_{\mathbb{1}}\\
			Z_e\\
			Z_m\\
			Z_f\\
		\end{pmatrix} =
		\begin{pmatrix}
			1 \\
			1\\
			0\\
			0
		\end{pmatrix} \text{ or } \begin{pmatrix}
		1\\
		0\\
		1\\
		0
	\end{pmatrix}, 
	\end{align} satisfy Eq.\ref{stv} with the $S$ and $T$ matrices from Eq.\ref{stz2}.
The two boundaries, with either $Z_{e}=1$ or $Z_m=1$,  correspond to the condensation of $e$ and $m$ excitations respectively. In fact, topological excitations on both gapped boundaries form a $\mathbb{Z}_2$ fusion ring. 

For example, when the $e$ excitation is condensed, it is identified with the trivial excitation, {\it i.e.,} $\mathbb{1}\sim e$. Also, since $e$ and $m$ excitations only differ by an $f$ excitation from the fusion rules, $m$ is identified with $f$. Therefore, there are only two distinct topological excitations on the boundary, namely $\{\mathbb{1},f\},$ which forms a $\mathbb{Z}_2$ fusion ring. The same thing happens when the $m$ excitation is condensed instead.

The \emph{phase transition} between these two $\mathbb{Z}_2$ symmetry-broken phases is an interesting question.\footnote{ From the boundary viewpoint, it is the critical point of the Ising model with a restricted Hilbert space.} Later, we will propose a similar question to gapped boundaries of double-Laughlin's states (see the paragraph around Eq.\ref{sameset}).
	\subsection{Local sector of a boundary}
	Notice that the above algorithm only works for bosonic topological orders, where the $S$ and $T$ matrices are well-defined. For fermionic topological orders, we constructed the bosonic extension, characterized by a bosonic $K$ matrix $K_B$. Finally, with the algorithm above, we worked through many examples and came up with the following recipe.
	
	Given a $K_F$, we can construct a bosonic extension on the top of $\mathbb{Z}_2$ topological order as in Sec.\ref{constuction}. We argue that this bosonic system must have an even number of gapped boundaries. These boundaries can be paired up, so that each pair is associated with a single gapped boundary of $K_F$. Conceptually, gapped boundaries of $K_B$ always show up in pair because they are associated with condensing $e$ and $m$ excitations of $\mathbb{Z}_2$ topological order respectively\footnote{Either $e$ or $m$ excitations \emph{MUST} be condensed in our construction because on the other side of the boundary is the vacuum. Therefore, the $\mathbb{Z}_2$ topological order must be destroyed in some ways on the boundary.}. 
	
	This is obviously an artifact of the construction, One can choose a different bosonic theory, {\it e.g.,} the Ising model, to start with and obtain a different set of redundancies. Our second claim is that such redundancies can be effectively removed by restricting to a maximal subset in which excitations have trivial mutual statistics with the electron. In other words, we can restrict the Hilbert space to the ``local sectors''.
	
	Concretely, assume that 
	\begin{align}\label{sb}
		S_B\equiv \{l_1, l_2,...,l_n\}
	\end{align} is a gapped boundary of $K_B$, where $l_i$'s are the condensed anyons on a boundary. Then, our claim says that the local sector would be a subset of $S_B$, named $S_{low}$, such that
	\begin{align}
		S_{low}=\{l\in S_B|\theta_{l,l_{ele}}= 0 \mod 2\pi\}.
	\end{align} In other words, excitations in $S_F$ are local with respect to the chosen electron $l_{ele}$. 

Also, in all examples we tried, this procedure always excludes half of the excitations in $S_B$, {\it i.e.}, \begin{align}\label{sbsf}
		|S_B| = 2\cdot |S_{low}|.
	\end{align} This should be intuitive if we think of a gapped boundary of $K_B$ as (very schematically) consisting of two parts\footnote{See the paragraph after Eq.\ref{stb} for a more precise statement and justification.}:
	\begin{align}\label{bdrysum}
		S_B = ``S_{low}+\text{trivial}" \cup ``S_{low}+ \text{condensed } e",
	\end{align} where $e$ is, as before, the $\mathbb{Z}_2$ charge. 

It is natural to expect there exists another gapped boundary that contains the same $S_F$ but with the condensation of $m$, {\it i.e.}, the $\mathbb{Z}_2$ flux. Put differently, the gapped boundaries of $K_B$ always come in pairs because the $\mathbb{Z}_2$ topological order has two gapped boundaries.
	
      	\section{Some examples}\label{example}
	
Here we present results for the $\nu=1-\frac{1}{m}$ fractional quantum Hall states.  The edges of these FQH states have zero chiral central
charge, and seems can be gapped.  However, it was pointed out by Levin
\cite{L13017355} that despite the  zero chiral central charge, most of those
quantum Hall states have no gapped edges.  

Our computation shows that, when $m\leq 50$,
only when $m$ is a square does there exist gapped boundaries. 
This is consistent with the result in Ref. \cite{L13017355}. Moreover, since the modular covariant
partition functions were found explicitly, our approach allows us to explicitly
find the anyons condensed on a boundary. 

In fact, these FQH states all have exactly one boundary, forming a simple, fusion ring structure. We expect this to hold in general (even for large $m$'s).

	\subsection{$\nu=1-\frac{1}{m}$ Laughlin's states}
	In this case, $m$ is an odd number, {\it i.e.,} $m\in 2\mathbb{Z}+1$.
	
	Take $\nu=1-\frac{1}{9}$ Laughlin's states for example. The $K$ matrix is given by 
	\begin{align}
		K_F=\begin{pmatrix}
			1 & 1 \\
			1 & -8 
		\end{pmatrix}.
	\end{align}
	The associated bosonic $K$ matrix is 
	\begin{align}
		K_B=\begin{pmatrix}
			0 & 2 & 1 & 0\\
			2 & 0 & 1 & 0\\
			1 & 1 & 2 & 1 \\
			0 & 0 & 1 & -8 
		\end{pmatrix}.
	\end{align}
	With our algorithm, there exist two gapped boundaries for $K_B$
	\begin{align}\label{sb1}
		S_B = \{ \begin{pmatrix}
			0\\
			0\\
			0\\
			0
		\end{pmatrix}, \begin{pmatrix}
			0\\
			-6\\
			0\\
			0
		\end{pmatrix}, \begin{pmatrix}
			0\\
			-12\\
			0\\
			0
		\end{pmatrix}, \begin{pmatrix}
			0\\
			-3\\
			0\\
			0
		\end{pmatrix}, \begin{pmatrix}
			0\\
			-9\\
			0\\
			0
		\end{pmatrix}, \begin{pmatrix}
			0\\
			-15\\
			0\\
			0
		\end{pmatrix}\}
	\end{align} and 
	\begin{align}\label{sb2}
		S_B' = \{ \begin{pmatrix}
			0\\
			0\\
			0\\
			0
		\end{pmatrix}, \begin{pmatrix}
			0\\
			-6\\
			0\\
			0
		\end{pmatrix}, \begin{pmatrix}
			0\\
			-12\\
			0\\
			0
		\end{pmatrix}, \begin{pmatrix}
			0\\
			-1\\
			0\\
			-1
		\end{pmatrix}, \begin{pmatrix}
			0\\
			-7\\
			0\\
			-1
		\end{pmatrix}, \begin{pmatrix}
			0\\
			-13\\
			0\\
			-1
		\end{pmatrix}\}.
	\end{align}
	
	It is straightforward to check that the local part, {\it i.e.,} having trivial mutual statistics with $l_{ele}$, is precisely the intersection of two sets, namely
	\begin{align}\label{set}
	S_{low}=\{ \begin{pmatrix}
			0\\
			0\\
			0\\
			0
		\end{pmatrix}, \begin{pmatrix}
			0\\
			-6\\
			0\\
			0
		\end{pmatrix}, \begin{pmatrix}
			0\\
			-12\\
			0\\
			0
		\end{pmatrix}\}.
	\end{align} Via the mapping Eq.\ref{mapping}, there exist unique (up to equivalence relation) excitations of the original fermionic system that are associated with Eq.\ref{set}, namely
	\begin{align}\label{bdry9}
		S_{F} = \{\begin{pmatrix}
			0\\
			0
		\end{pmatrix}, \begin{pmatrix}
			3\\
			0
		\end{pmatrix}, \begin{pmatrix}
			6\\
			0
		\end{pmatrix}\}
	\end{align}We conclude by claiming that they are the anyons that has to be condensed to form the gapped boundary of the $\nu=1-\frac{1}{9}$ FQH state.

	Similarly, for the $\nu=1-\frac{1}{25}$ FQH state, the $K$ matrix is
	\begin{align}
		K_F = \begin{pmatrix}
			1 & 1\\
			1 & -24
		\end{pmatrix}
	\end{align} and 
	\begin{align}
		K_B=\begin{pmatrix}
			0 & 2 & 1 & 0\\
			2 & 0 & 1 & 0\\
			1 & 1 & 2 & 1 \\
			0 & 0 & 1 & -24
		\end{pmatrix}.
	\end{align}
	
	Again there exist two gapped boundaries of $K_B$,
	\begin{align}
		S_B=\{&\begin{pmatrix}
			0\\
			0\\
			0\\
			0
		\end{pmatrix},\begin{pmatrix}
			0\\
			-10\\
			0\\
			0
		\end{pmatrix},\begin{pmatrix}
			0\\
			-20\\
			0\\
			0
		\end{pmatrix},\begin{pmatrix}
			0\\
			-30\\
			0\\
			0
		\end{pmatrix},\begin{pmatrix}
			0\\
			-40\\
			0\\
			0
		\end{pmatrix},\nonumber\\
		&\begin{pmatrix}
			0\\
			-5\\
			0\\
			0
		\end{pmatrix},\begin{pmatrix}
			0\\
			-15\\
			0\\
			0
		\end{pmatrix},\begin{pmatrix}
			0\\
			-25\\
			0\\
			0
		\end{pmatrix},\begin{pmatrix}
			0\\
			-35\\
			0\\
			0
		\end{pmatrix},\begin{pmatrix}
			0\\
			-45\\
			0\\
			0
		\end{pmatrix}\} 
	\end{align} and 
	\begin{align}
		S_B'=\{&\begin{pmatrix}
			0\\
			0\\
			0\\
			0
		\end{pmatrix},\begin{pmatrix}
			0\\
			-10\\
			0\\
			0
		\end{pmatrix},\begin{pmatrix}
			0\\
			-20\\
			0\\
			0
		\end{pmatrix},\begin{pmatrix}
			0\\
			-30\\
			0\\
			0
		\end{pmatrix},\begin{pmatrix}
			0\\
			-40\\
			0\\
			0
		\end{pmatrix},\nonumber\\
		&\begin{pmatrix}
			0\\
			-3\\
			0\\
			0
		\end{pmatrix},\begin{pmatrix}
			0\\
			-13\\
			0\\
			0
		\end{pmatrix},\begin{pmatrix}
			0\\
			-23\\
			0\\
			0
		\end{pmatrix},\begin{pmatrix}
			0\\
			-33\\
			0\\
			0
		\end{pmatrix},\begin{pmatrix}
			0\\
			-43\\
			0\\
			0
		\end{pmatrix}\} .
	\end{align}
	
	Again, there exists only one low-energy gapped boundary,
	\begin{align}
	S_{low}=	\{&\begin{pmatrix}
			0\\
			0\\
			0\\
			0
		\end{pmatrix},\begin{pmatrix}
			0\\
			-10\\
			0\\
			0
		\end{pmatrix},\begin{pmatrix}
			0\\
			-20\\
			0\\
			0
		\end{pmatrix},\begin{pmatrix}
			0\\
			-30\\
			0\\
			0
		\end{pmatrix},\begin{pmatrix}
			0\\
			-40\\
			0\\
			0
		\end{pmatrix}\}.
	\end{align}
	The associated $l_F$'s are
	\begin{align}
	S_F =	\{
		\begin{pmatrix}
			0\\
			0
		\end{pmatrix},\begin{pmatrix}
			5\\
			0
		\end{pmatrix},\begin{pmatrix}
			10\\
			0
		\end{pmatrix},\begin{pmatrix}
			15\\
			0
		\end{pmatrix},\begin{pmatrix}
			20\\
			0
		\end{pmatrix}\}.
	\end{align}
	\subsection{$m\in 2 \mathbb{Z} $}
	For completeness, let's consider the cases where $m$ is an even number. Notice that these are \emph{NOT} the non-abelian FQH states seen in the experiments. 
	
	Since we only want one odd number on the diagonal, the fermionic $K$ matrix reads
	\begin{align}
		K_F=\begin{pmatrix}
			1 & 0\\
			0 & -m
		\end{pmatrix}, \quad m\in 2\mathbb{Z},
	\end{align} and thus the bosonic extension reads
\begin{align}
	K_B = \begin{pmatrix}
		0 & 2 & 1 & 0\\
		2 & 0 & 1 & 0\\
		1 & 1 & 2 & 0\\
		0 & 0 & 0 & -m
	\end{pmatrix}.
\end{align}

Our algorithm shows that, similar to the case of Laughlin's states, there exists exactly one gapped boundary when $m$ is a square. For example, when $m=4$, the bosonic extension has two gapped boundaries,
\begin{align}
	S_B=\{\begin{pmatrix}
		0 \\ 0 \\ 0 \\ 0
		\end{pmatrix}, \begin{pmatrix}
		0 \\ 2 \\ 0 \\ -2
	\end{pmatrix}, \begin{pmatrix}
	0 \\ 3 \\ 0 \\ -1
\end{pmatrix}, \begin{pmatrix}
	0 \\ 1 \\ 0 \\ -3
\end{pmatrix}\}
\end{align} and 
\begin{align}
	S_B'=\{\begin{pmatrix}
		0 \\ 0 \\ 0 \\ 0
	\end{pmatrix}, \begin{pmatrix}
		0 \\ 2 \\ 0 \\ -2
	\end{pmatrix}, \begin{pmatrix}
		0 \\ 1 \\ 0 \\ -1
	\end{pmatrix}, \begin{pmatrix}
		0 \\ 3 \\ 0 \\ -3
	\end{pmatrix}\}.
\end{align} 

The low energy part is 
\begin{align}
S_{low}=\{\begin{pmatrix}
	0 \\ 0 \\ 0 \\ 0
\end{pmatrix}, \begin{pmatrix}
	0 \\ 2 \\ 0 \\ -2
\end{pmatrix}\},
\end{align} which associated $l_F$'s are
\begin{align}
S_{F}=\{\begin{pmatrix}
	0 \\ 0
	\end{pmatrix}, \begin{pmatrix}
	-1 \\ -2
	\end{pmatrix}\}.
\end{align} We claim that is the condensation on the gapped boundary of $K_F=\begin{pmatrix}
1 &0\\
0 & -4
\end{pmatrix}$ FQH states.

Similarly, we found that, when $m=16,$ the gapped boundary is the condensation of
\begin{align}
	\{\begin{pmatrix}
		0 \\ 0
	\end{pmatrix}, \begin{pmatrix}
		 0 \\ -8
	\end{pmatrix},\begin{pmatrix}
	-1 \\ -4
\end{pmatrix}, \begin{pmatrix}
-1\\-12
\end{pmatrix}\}.
\end{align}

	\subsection{Stacking of Laughlin's states} We can also consider the stacking of $\nu=\frac{1}{m}$ and $\nu=\frac{1}{n}$ Laughlin's states. An interesting finding is that the double-Laughlin's states, which $K$ matrix is
	\begin{align}
		K_F=\begin{pmatrix}
			m & m \\
			m & 0 
		\end{pmatrix}, \quad m\in 2\mathbb{Z}+1,
	\end{align} has exactly $2$ gapped boundaries when $m$ is a prime number (See Table.\ref{tab:num}). 
	
	As an example, when $m=1$, {\it i.e.},
	\begin{align}
		K_F=\begin{pmatrix}
			3 &3\\
			3 & 0
		\end{pmatrix},
	\end{align} the four gapped boundaries of $K_B$ are
\begin{align}
	S_{B,1}=\{\begin{pmatrix}
		0\\0\\0\\0
	\end{pmatrix},\begin{pmatrix}
	0\\-2\\0\\0
\end{pmatrix},\begin{pmatrix}
0\\-4\\0\\0
\end{pmatrix},\begin{pmatrix}
0\\-1\\0\\0
\end{pmatrix},\begin{pmatrix}
0\\-3\\0\\0
\end{pmatrix},\begin{pmatrix}
0\\-5\\0\\0
\end{pmatrix}\}\\	
S_{B,2}=\{\begin{pmatrix}
0\\0\\0\\0
\end{pmatrix},\begin{pmatrix}
0\\-2\\0\\0
\end{pmatrix},\begin{pmatrix}
0\\-4\\0\\0
\end{pmatrix},\begin{pmatrix}
0\\1\\0\\-3
\end{pmatrix},\begin{pmatrix}
0\\3\\0\\-3
\end{pmatrix},\begin{pmatrix}
0\\5\\0\\-3
\end{pmatrix}\}\\
S_{B,3}=\{\begin{pmatrix}
	0\\0\\0\\0
\end{pmatrix},\begin{pmatrix}
	0\\2\\0\\-2
\end{pmatrix},\begin{pmatrix}
	0\\4\\0\\-4
\end{pmatrix},\begin{pmatrix}
	0\\-1\\0\\-2
\end{pmatrix},\begin{pmatrix}
	0\\-3\\0\\-0
\end{pmatrix},\begin{pmatrix}
	0\\7\\0\\-4
\end{pmatrix}\}\\
S_{B,4}=\{\begin{pmatrix}
	0\\0\\0\\0
\end{pmatrix},\begin{pmatrix}
	0\\2\\0\\-2
\end{pmatrix},\begin{pmatrix}
	0\\4\\0\\-4
\end{pmatrix},\begin{pmatrix}
	0\\-1\\0\\-1
\end{pmatrix},\begin{pmatrix}
	0\\3\\0\\-3
\end{pmatrix},\begin{pmatrix}
	0\\5\\0\\-5
\end{pmatrix}\}.
\end{align}

Clearly, they can be paired up, and the intersections precisely give us two distinct low-energy boundaries\footnote{As a reminder, Eq.\ref{slow1} and \ref{slow2} are indeed low energy because each of the anyon in the sets has their first two gauge charges summing up to an even number.},
\begin{align}\label{slow1}
	S_{low,1}=S_{B,1}\cap S_{B,2}=\{\begin{pmatrix}
		0\\0\\0\\0
	\end{pmatrix},\begin{pmatrix}
		0\\-2\\0\\0
	\end{pmatrix},\begin{pmatrix}
		0\\-4\\0\\0
	\end{pmatrix}\}\\
\label{slow2}S_{low,2}=S_{B,3}\cap S_{B,4}=\{\begin{pmatrix}
	0\\0\\0\\0
\end{pmatrix},\begin{pmatrix}
	0\\2\\0\\-2
\end{pmatrix},\begin{pmatrix}
	0\\4\\0\\-4
\end{pmatrix}\}.
\end{align}
	Again, via the mapping, the gapped boundaries of the original fermionic system are labeled by the low-energy condensations (in the same sense of Eq.\ref{bdry9})
	\begin{align}\label{sf12}
		S_{F,1}& = \{\begin{pmatrix}
			0\\0
		\end{pmatrix},\begin{pmatrix}
			1\\0
		\end{pmatrix},\begin{pmatrix}
			2\\0
		\end{pmatrix}\}\\
		S_{F,2}& = \{\begin{pmatrix}
			0\\0
		\end{pmatrix},\begin{pmatrix}
			-1\\-2
		\end{pmatrix}, \begin{pmatrix}
			-2\\-4
		\end{pmatrix}\}.
	\end{align}
	
	\begin{table}[]
		\centering
		\begin{tabular}{|c|c|}
			\hline
			$m$ & \# \\
			\hline
			1 & 1\\
			3 & 2\\
			5 & 2\\
			7 & 2\\
			9 & 3\\
			11 & 2\\
			13 & 2\\
			15 & 4\\
			17 & 2\\
			19 & 2\\
			21 & 4\\
			23 & 2\\
			25 & 3\\
			27 & 4\\
			29 & 2\\
			31 & 2\\
			49 & 3\\
			\hline
		\end{tabular}
		\caption{The number of gapped boundaries of double-$\nu=\frac{1}{m}$ Laughlin's states. When $m$ is a prime, the number is $2$.}
		\label{tab:num}
	\end{table}

More generally, consider the stacking of two different Laughlin's states, say $\nu=\frac{1}{m}$ and $\nu=\frac{1}{n}$ with $m,n\in2\mathbb{Z}+1$, {\it i.e.,}
\begin{align}
	K_F=\begin{pmatrix}
		m &m\\
		m & m-n
	\end{pmatrix}.
\end{align} The numerical results are collected in Table.\ref{tab:stack}. Looking at the first three columns, we observe that the system allows gapped boundaries when the product $mn$ is a square. In other words, if $m$ is not a square, then $n$ equals to $m$ multiplied by an odd square, {\it i.e.,} $n=mk^2,$ $k\in 2\mathbb{Z}+1$; if $m$ is a square, then $n$ itself must be a square as well.
	\begin{table}[]
	\centering
	\begin{tabular}{|c|c|c|c|}
		\hline
		$m$ & $n$ & \# & boundary fusion ring\\
		\hline
		1 & 1 & 1& \\
		\hline
		1 & 9 &1& $\mathbb{Z}_2\times \mathbb{Z}_3$\\
		\hline
		1 & 25 & 1& $\mathbb{Z}_2\times \mathbb{Z}_5$\\
		\hline
		1&49&1& $\mathbb{Z}_2\times \mathbb{Z}_7$\\
		\hline
		1&81&1&\\ 
		\hline
		\hline
		3 & 3 &2& $\mathbb{Z}_2\times \mathbb{Z}_3$, $\mathbb{Z}_2\times \mathbb{Z}_3$\\
		\hline
		3&27 & 2 & $\mathbb{Z}_2\times \mathbb{Z}_9$, $\mathbb{Z}_2\times \mathbb{Z}_9$\\
		\hline
		3 & 75 & 2&\\
		\hline
		\hline 
		5 & 5 & 2& $\mathbb{Z}_2\times \mathbb{Z}_5$, $\mathbb{Z}_2\times \mathbb{Z}_5$\\
		\hline
		5 &  45 & 2&\\
		\hline
		\hline 
		7 & 7 &2&$\mathbb{Z}_2\times \mathbb{Z}_7$, $\mathbb{Z}_2\times \mathbb{Z}_7$ \\
		\hline
		7 & 63 & 2&\\
		\hline
		\hline
		9 & 1 &1&$\mathbb{Z}_2\times \mathbb{Z}_3$\\
		\hline
		9 & 9 &3&\\
		\hline
		9 & 25 & 1&\\
		\hline
		9&49&1&\\
		\hline
		9&81&3&\\
		\hline
	\end{tabular}
	\caption{The number of gapped boundaries of the stacking of $\nu=\frac{1}{m}$ and $\nu=-\frac{1}{n}$ Laughlin's states, where $m$ ranges from $1$ to $9$ and $n$ ranges from $1$ to $101$. Both $m$ and $n$ are odd. We observe that the system allows gapped boundaries when $mn$ is a square. We computed several boundary fusion rings and listed them in the forth column. }
	\label{tab:stack}
\end{table}
	
	As an example, when $m=3$ and $n=27$, there exists two types of low-energy gapped boundaries:
	\begin{align}\label{slow1stack}
		S_{low,1}=&\{\begin{pmatrix}
			0\\
			0\\
			0\\
			0
		\end{pmatrix},\begin{pmatrix}
		0\\-18\\ 0\\0
	\end{pmatrix},\begin{pmatrix}
	0\\-36\\0\\0
\end{pmatrix},\begin{pmatrix}
0\\-2\\0\\-2
\end{pmatrix},\begin{pmatrix}
0\\-20\\0\\-2
\end{pmatrix},\nonumber\\
 &\begin{pmatrix}
0\\-38\\0\\-2
\end{pmatrix},\begin{pmatrix}
0\\-4\\0\\-4d
\end{pmatrix},\begin{pmatrix}
0\\-22\\0\\-4
\end{pmatrix},\begin{pmatrix}
0\\-40\\0\\-2
\end{pmatrix}\}\\
\end{align}
and 
\begin{align}\label{slow3stack}
S_{low,2}=&\{\begin{pmatrix}
	0\\
	0\\
	0\\
	0
\end{pmatrix},\begin{pmatrix}
	0\\-18\\ 0\\0
\end{pmatrix},\begin{pmatrix}
	0\\-36\\0\\0
\end{pmatrix},\begin{pmatrix}
	0\\-8\\0\\-2
\end{pmatrix},\begin{pmatrix}
	0\\-26\\0\\-2
\end{pmatrix},\nonumber\\
&\begin{pmatrix}
	0\\-44\\0\\-2
\end{pmatrix},\begin{pmatrix}
	0\\2\\0\\-4
\end{pmatrix},\begin{pmatrix}
	0\\-16\\0\\-4
\end{pmatrix},\begin{pmatrix}
	0\\-34\\0\\-4
\end{pmatrix}\}.
	\end{align}

	\subsection{Boundary topological excitations}
	Our construction allows us to find the boundary topological excitations explicitly. For example, for $\nu=1-\frac{1}{9}$ Laughlin's state, the gapped boundary condenses excitations in Eq.\ref{bdry9}. These new condensations give rise to new equivalence relations on the boundary, in addition to the columns of $K_F$.
	
	Explicitly, the equivalence relations of the boundary excitations become
	\begin{align}
		S_{triv.}=\{\begin{pmatrix}
			3\\0
		\end{pmatrix},
	\begin{pmatrix}
	1\\1
	\end{pmatrix},
\begin{pmatrix}
	1\\-8
\end{pmatrix}\}
	\end{align} such that $l_{F,bdry}\sim l_{F,bdry}+l_{triv.}$, where $l_{triv.}\in S_{triv}$ and $l_{F,bdry}$ are boundary excitations. Then it is straightforward to check that, if the electron is taken to be trivial, there are three distinct boundary excitations,
	\begin{align}
		S_{F,bdry}=\{\begin{pmatrix}
			0\\0
		\end{pmatrix},\begin{pmatrix}
			1\\0
		\end{pmatrix},\begin{pmatrix}
			2\\0
		\end{pmatrix}\}.
	\end{align} All together, they form a $\mathbb{Z}_3$ fusion ring. 

Nevertheless, we know that the electron never condenses; it is merely a mathematical convenience to treat it as a trivial excitation. If the electron, namely $(1 , 1)^T$ in this case, is added back, then all six of them will form a $\mathbb{Z}_3\times \mathbb{Z}_2=\mathbb{Z}_6$ fusion ring. 

When $m$ is even, the same story occurs. For example, when $m=4,$ one can show that the boundary excitations are
\begin{align}
	S_{F,bdry}=\{\begin{pmatrix}
		0\\0
	\end{pmatrix},\begin{pmatrix}
	0\\1
\end{pmatrix}\}.
\end{align} These excitations, along with the addition of electron, form a $\mathbb{Z}_2\times\mathbb{Z}_2$ fusion ring, as expected.

We believe that this is general. For $\nu=1-\frac{1}{m}$ FQH states, where $m=k^2$, the boundary topological excitations form a $\mathbb{Z}_2\times\mathbb{Z}_k$ fusion ring, where $\mathbb{Z}_2$ comes from the addition of an electron.
	
	Of course, a similar procedure can be carried out for double-Laughlin's states. Consider $K_F=\begin{pmatrix}
		3 &3\\
		3 &0
	\end{pmatrix}$ and condense either $S_{F,1}$ or $S_{F,2}$ in Eq.\ref{sf12}. Perhaps surprisingly, we again obtain a $\mathbb{Z}_6$ fusion ring for both gapped boundaries; in particular, even the excitations themselves are the same,
	\begin{align}\label{sameset}
		\{\begin{pmatrix}
			0\\0
		\end{pmatrix},\begin{pmatrix}
			1\\0
		\end{pmatrix},\begin{pmatrix}
			2\\0
		\end{pmatrix}\}.
	\end{align} It may be interesting to ask whether there is a \emph{phase transition} between them.

The situation becomes more complicated when $m$ and $n$ are different. See the forth column of Tab.\ref{tab:stack}. Notice that we only have partial results for the third column, since we do not find a good way to identify independent equivalence relations on the boundary. We managed to solve the $(m,n)=(3,27)$ case. The gapped boundaries of $K_F$ can be obtained from Eq.\ref{slow1stack}, \ref{slow3stack}, and \ref{mapping},
\begin{align}\label{sf1}
	S_{F,1}=&\{\begin{pmatrix}
		0\\0
	\end{pmatrix},\begin{pmatrix}
	9\\0
\end{pmatrix},\begin{pmatrix}
18\\0
\end{pmatrix},\begin{pmatrix}
1\\-2
\end{pmatrix},\begin{pmatrix}
10\\-2
\end{pmatrix},\nonumber\\
&\begin{pmatrix}
19\\-2 \end{pmatrix}, \begin{pmatrix}
2\\-4
\end{pmatrix}, \begin{pmatrix}
11\\-4
\end{pmatrix},\begin{pmatrix}
20\\-4
\end{pmatrix}\}
\end{align}

and \begin{align}\label{sf2}
	S_{F,2}=&\{\begin{pmatrix}
	0\\0
\end{pmatrix},\begin{pmatrix}
	9\\0
\end{pmatrix},\begin{pmatrix}
	18\\0
\end{pmatrix},\begin{pmatrix}
	4\\-2
\end{pmatrix},\begin{pmatrix}
	13\\-2
\end{pmatrix},\nonumber\\
&\begin{pmatrix}
22\\-2
\end{pmatrix},\begin{pmatrix}
-1\\-4
\end{pmatrix},\begin{pmatrix}
8\\-4
\end{pmatrix},\begin{pmatrix}
17\\-4
\end{pmatrix}\}.
\end{align}

The equivalence relations on each boundary are
\begin{align}
	S_{triv.,1}=\{\begin{pmatrix}
	 9\\0
	\end{pmatrix},\begin{pmatrix}
	1\\-2
\end{pmatrix}\}\cup \{K_{F,i}\}\\
	S_{triv.,2}=\{\begin{pmatrix}
	9\\0
\end{pmatrix},\begin{pmatrix}
	4\\-2
\end{pmatrix}\}\cup \{K_{F,i}\},
\end{align} where $\{K_{F,i}\}=\{\begin{pmatrix}
3\\3
\end{pmatrix},\begin{pmatrix}
3\\-24
\end{pmatrix}\}$. By inspection, it happens that the inequivalent boundary excitations are
\begin{align}\label{mark}
	S_{F,bdry,1}=S_{F,bdry,2}=\{\begin{pmatrix}
		0\\0
	\end{pmatrix},\begin{pmatrix}
	0\\1
\end{pmatrix},\cdots, \begin{pmatrix}
0\\8
\end{pmatrix}\}.
\end{align} In other words, the boundary excitations of both boundaries form a $\mathbb{Z}_2\times \mathbb{Z}_9$ fusion ring.

We did not fully determine the boundary excitations of all $(m,n)$ cases, {\it i.e.,} empty slots in Table.\ref{tab:stack}. The difficulty boils down to: given the equivalence relations that are not linearly independent ({\it e.g.,} Eq.\ref{sf1} or \ref{sf2}), is there a canonical way to identify all distinct vectors up to equivalence relations? We did not find a good solution to this question, but, based on the partial results, we suspect that  when $mn=k^2$ for some odd number $k$, there exists gapped boundaries, whose topological excitations always form a $\mathbb{Z}_2\times \mathbb{Z}_k$ fusion ring. 
	
	\section{An alternative perspective}\label{alter}
	We have described an approach that combines the philosophy of effective field theory and an algebraic algorithm to study the gapped boundaries of $2+1$D  abelian fermionic topological orders. However, the argument of the former may be suspicious to some readers. Is it really okay to simply throw away the non-local anyons in Eq.\ref{sb} and claim the rest to constitute a gapped boundary of $K_F$? 
	
	Here we present an alternative perspective that confirms the results in the previous section, hopefully boosting the readers' confidence in our $K_B$ construction.
	As commented earlier, the condensation of non-local anyons is required to break the $\mathbb{Z}_2$ topological order into the vacuum, which by usual definition has trivial topological order. Nevertheless, our philosophy is that the universe may secretly emerge from a topologically-ordered matter at very high $UV$ scale, which is why we can view the electron as an emergent excitation in the first place. 
	
	In other words, in our $K_B$ construction we should let ``the vacuum'' to have the $\mathbb{Z}_2$ topological order. The effective field theory is consequently described by a new $K$ matrix,
	\begin{align}\label{newk}
		K = \begin{pmatrix}
			K_{\mathbb{Z}_2} & 0\\
			0 & K_B
		\end{pmatrix},
	\end{align} where $K_B$ is defined as previously. We then try to find the gapped boundaries of this new $K$ matrix.
	
	Since experimentally we never observed objects of such underlying topological order, {\it e.g.,} defects, $e$- and $m$-like particles in the Standard Model, etc, we must require the interface between the vacuum and anything to be a \emph{trivial} boundary. What does \emph{trivial} mean? Intuitively, it means that when an excitation passes through a trivial boundary, nothing should happen; for example, an $e$ particle in the vacuum should remain being type-$e$ when crossing a vacuum-material interface. Therefore, we suggest that a trivial gapped boundary requires the condensation of $\mathbb{1}-\mathbb{1}$, $e-e$, $m-m$, and $f-f$ (see Appx.\ref{bdryz2z2} for more details).
	
	We computed the gapped boundaries of our new $K$ matrix for the $\nu=1-\frac{1}{9}$ FQH state and found a unique solution that satisfies the requirement,
	\begin{align}
		S_{t.b.} = \{S_\mathbb{1},S_e, S_m, S_f \},
	\end{align} where
	\begin{align}\label{stb}
		S_\mathbb{1} & = \{ \begin{pmatrix}
			0\\
			0\\
			0\\
			0\\
			0\\
			0
		\end{pmatrix}, \begin{pmatrix}
			0\\
			0\\
			0\\
			-6\\
			0\\
			0\\
		\end{pmatrix}, \begin{pmatrix}
			0\\
			0\\
			0\\
			-12\\
			0\\
			0\\
		\end{pmatrix} \} \\
		S_e & = \{ \begin{pmatrix}
			1\\
			0\\
			0\\
			-1\\
			0\\
			-1
		\end{pmatrix}, \begin{pmatrix}
			1\\
			0\\
			0\\
			-7\\
			0\\
			-1\\
		\end{pmatrix}, \begin{pmatrix}
			1\\
			0\\
			0\\
			-13\\
			0\\
			-1\\
		\end{pmatrix} \} 
		\end{align}
\begin{align}
		S_m & = \{ \begin{pmatrix}
			0\\
			1\\
			0\\
			-3\\
			0\\
			0
		\end{pmatrix}, \begin{pmatrix}
			0\\
			1\\
			0\\
			-9\\
			0\\
			0\\
		\end{pmatrix}, \begin{pmatrix}
			0\\
			1\\
			0\\
			-15\\
			0\\
			0\\
		\end{pmatrix} \} 
			\end{align}
\begin{align}
		S_f & = \{ \begin{pmatrix}
			1\\
			1\\
			0\\
			2\\
			0\\
			-1
		\end{pmatrix}, \begin{pmatrix}
			1\\
			1\\
			0\\
			-4\\
			0\\
			-1\\
		\end{pmatrix}, \begin{pmatrix}
			1\\
			1\\
			0\\
			-10\\
			0\\
			-1\\
		\end{pmatrix} \}.
	\end{align} The subscript of $S_i$ indicates the $i-i$ bound states of the underlying $\mathbb{Z}_2$ topological order; for example, the first two entries of $S_e$ are $(1,0)^T$, {\it i.e.}, the $e$ excitation of vacuum. 
	
Now, we can view this as a more precise notion of Eq.\ref{bdrysum}. For example, looking at the last four entries of $S_\mathbb{1}$ and $S_e$, we may argue that bounding with an $e$ particle of $K_B$ amounts to the addition of gauge charges $(0,-1,0,-1)^T$. Similarly, one can also argue that $(0,-3,0,0)^T$ is the $m$ particle. Indeed, they are both bosons and have $\pi-$mutual statistics.
	
	In other words, this justifies our interpretation of Eq.\ref{sb1}, \ref{sb2}, and especially \ref{set}. The bosonic extension of the $\nu=1-\frac{1}{9}$ FQH state has two gapped boundaries, associated with $e$ and $m$ particles respectively,
	\begin{align}
		S_B ``\approx" S_{\mathbb{1}}\cup S_m
	\end{align} and 
	\begin{align}
		S_B' ``\approx" S_{\mathbb{1}} \cup  S_e,
	\end{align} while the low-energy part is simply
	\begin{align}
		S_{low} ``\approx" S_{\mathbb{1}}.
	\end{align} The $``\approx"$ here means ignoring the vacuum, {\it i.e.}, ignoring the first two gauge charges of each excitation.
	
	In this alternative construction, we avoid the potential doubt on restricting to the local sector. Instead, we invoke the notion of the \emph{trivial boundary}, which may be more convincing to some readers. However, an obvious disadvantage of this approach is that the number of excitations gets multiplied by four. With our algorithm, it takes significantly more time to find the gapped boundaries of such larger $K$ matrix. Therefore, we treat this perspective as a \emph{justification} to our previous approach, which was motivated by the philosophy of effective field theory.
	
	\section{Discussion}
	In this work, we developed a systematic approach to study gapped boundaries of $2+1$D abelian FQH states. In particular, we devised an algorithm to find the anyon condensation for gapping the boundaries. The main trick we introduced is to find a bosonic extension of a fermionic topological order. This way, the equivalence relations of a fermionic topological order become the low-energy part of those of the bosonic extension. 
	
	The correspondence can be constructed rather straightforwardly for the abelian case with the help of the $K$ matrix, as shown in this work. For the non-abelian case, it is not clear how to establish the correspondence explicitly, but we believe that the same framework would work: 
	\begin{mdframed}
		\begin{itemize}
			\vspace{5 pt}
			\item Find a bosonic extension and its equivalence relations.
			\item Then identify the electron and thus the low-energy/local part of the equivalence relations. 
			\vspace{3 pt}
		\end{itemize}
	\end{mdframed} Hopefully this will provide a systematic construction of fermionic topological orders. We will pursue this direction in the future work.
	
	We end the piece by mentioning some other possible future directions.
	\begin{enumerate}
		\item \underline{Dynamics of gapping the boundary}: Since the anyon condensation on the gapped boundaries can be found explicitly, this may provide an opportunity to work out the \emph{dynamics} as well: for example, the real-time formation of gapped boundaries. One may try starting with the gapless boundary theory
		\begin{align}
			\mathcal{L}=\frac{K_{IJ}}{4\pi}\partial_x \phi_I \partial_t \phi_J-\frac{V_{IJ}}{4\pi}\partial_x \phi_I\partial_x\phi_J.
		\end{align} along with a Higgs term
		\begin{align}
			\mathcal{L}_{\text{Higgs}}\approx A_I \cos( B_I K_{IJ}\phi_J),
		\end{align} where $A_I$ and $B_I$ are real parameters. It is interesting to, first, find a precise form of the Higgs term and, second, try to solve the theory.
		
		\item \underline{Phase transition between gapped boundaries}: Our result shows that different gapped boundaries, {\it i.e.}, different anyon condensation, may give rise to the same boundary excitations (see the paragraphs near Eq.\ref{sameset}). It is interesting to ask whether there could be a phase transition between them from the boundary point of view. Of course, this may be a very difficult direction, since one may need to work out the precise Lagrangian of the boundary theory first.
		
		\item \underline{A purely $K_F$ algorithm?:} Our $K_B$ construction gives the boundary excitations explicitly. Then it is natural to ask if we can devise another algorithm that only involves $K_F$. Perhaps equivalently, given the modular data of a fermionic system, can we tell whether the boundary excitations form a $\mathbb{Z}_N$ fusion ring or something more complicated, without resorting to the bosonic extension? 
		
		\item \underline{Gapless boundaries:} For starter, one can try to study the gapless boundaries from rational conformal field theories. In this case the modular covariant partition functions can be written as a function of the characters. The $S$ and $T$ matrices also have to change accordingly\cite{PhysRevResearch.1.033054}. However, before worrying about the subtleties of fermions, one may want to understand better the bosonic cases first.
		
	\end{enumerate}

	\section{Acknowledgement}

	Chang-Han Chen is supported by the Undergraduate Research Opportunity
Program (UROP) at MIT.  XGW is partially supported by NSF DMR-2022428 and by
the Simons Collaboration on Ultra-Quantum Matter, which is a grant from the
Simons Foundation (651446, XGW)

\appendix 
	\section{Hierarchical construction}\label{hiera}
	The idea of hierarchical construction is that when the density of an excitation reaches certain value, it will condense and form a new FQH state. The excitations of the new state will move in an effective ``magnetic field" (or more formally, the gauge fluxes) resulting from all the condensates and condense when its density reaches another certain value. In other words, the current of the $I$th condensate satisfies 
	\begin{align}
		j_I^\mu = \frac{1}{2\pi}\epsilon^{\mu\alpha\beta}\partial_\alpha a_{I\beta}.
	\end{align} 
	Since the same procedure can be carried out infinitely many times, forming condensates on top of each other, this construction was named \emph{hierarchical construction}.
	
	Concretely, let $K^{(n-1)}$ be an $(n-1)\times(n-1)$ $K$ matrix and the associated gauge fields be $a_{I\mu}$, $I\in{1,2,...,n-1}$. Assume that an excitation, whose gauge charges are labeled by $l\in \mathbb{Z}^{n-1}$, is now condensed. The new system will then be described by 
	\begin{align}
		K^{(n)}=\begin{pmatrix}
			K^{(n-1)} & -l\\
			-l^T & p_n
		\end{pmatrix},
	\end{align}
	where $p_n$ is an even number.

	A classic example is to construct $\nu=\frac{2}{5}$ FQH states from $\nu=\frac{1}{3}$ Laughlin's state. In this case, we start with the $K^{(1)}=\begin{pmatrix}
		3
	\end{pmatrix}$ and then condense the excitation with one unit of the gauge charge. The resulting $K$ matrix is 
	\begin{align}
		K^{(2)} = \begin{pmatrix}
			3 & -1\\
			-1 & 2
		\end{pmatrix},
	\end{align} and the charge vector is $q=(1,0)^T$. One can then verify that this indeed gives you $\nu=\frac{2}{5}$ from Eq.\ref{nu}. With the same procedure, one can check that 
	\begin{align}
		K = \begin{pmatrix}
			3 & -1 & 0 \\
			-1 & 2 & -1 \\
			0 & -1 & 2
		\end{pmatrix}
	\end{align} corresponds to $\nu = \frac{3}{7}$ FQH state.
	
	As mentioned in Sec.\ref{remarks}, $|\det K|$ can be identified with the volume of the fundamental parallelepiped. It is a fact that the number of anyons, as well as the ground state degeneracy (GSD), equals to $|\det K|^g$, where $g$ is the genus of the Riemann surface. On the torus, $g$ equals to one, and GSD$=|\det K|$, as expected. Therefore, $\nu=\frac{1}{3}$ Laughlin's state supports three types of anyons, while $\nu=\frac{2}{5}$ state supports five types.
	
	Another example that will be useful for us is the $K$ matrix of the $\mathbb{Z}_2$ topological order
	\begin{align}
		K=\begin{pmatrix}
			0 & 2 \\
			2& 0
		\end{pmatrix}.
	\end{align} This theory has $4$ types of anyons, namely
	\begin{align}
		\mathbb{1}=\begin{pmatrix}
			0\\
			0
		\end{pmatrix}, e = \begin{pmatrix}
			1\\
			0
		\end{pmatrix}, m = \begin{pmatrix}
			0\\
			1
		\end{pmatrix}, f = \begin{pmatrix}
			1\\
			1
		\end{pmatrix}.
	\end{align} $\mathbb{1}$ is a trivial excitation, while $e$ and $m$ are bosons and $f$ is a fermion. Interestingly, despite their self-statistics, $e,$ $m,$ and $f$ have $\pi$ mutual statistics with each other, {\it i.e.}, they are mutual semions. They also satisfy the fusion rule:
	\begin{align}
		&e\times m = f, e\times f =m, m\times f = e,\\
		&e\times e=m\times m = f\times f =\mathbb{1}
	\end{align} In the main text, we treat the $f$-type anyon as the electron to build fermionic topological orders.
	\section{Proofs of some statements}\label{proof}
	\subsection{$l$ must be an integer vector}\label{proof1}
	The mutual statistics of an excitation $l$ and a column of $K$ can be written as 
	\begin{align}
		\theta = 2\pi l^T K^{-1}K_i.
	\end{align}
	Notice that 
	\begin{align}
		\sum_j K^{-1}_{i,j}K_{j,l}=\delta_{i,l},
	\end{align} so
	\begin{align}
		&\theta = 2\pi l^T\cdot \mathbf{e_i}\nonumber\\
		&=2\pi l_i,
	\end{align} where $\mathbf{e_i}$ is the $i$-th unit vector. Having trivial mutual statistics requires $\theta$ to be multiples of $2\pi$. Hence,
	\begin{align}
		l_i\in \mathbb{Z},\quad \forall i=1,2,...,n
	\end{align}
	\subsection{Self- and mutual statistics of $l_{ele}$}\label{proof2}
	Given $K_B$ and $l_{ele}$,
	\begin{align}
		&K_B \cdot l_{ele} = \begin{pmatrix}
			0 & 2 & 1 &0  &...\\
			2 & 0 & 1 &0  &...\\
			1 & 1 & (1+K_F{}_{1,1}) & K_F{}_{1,2} &...\\
			0 & 0 & K_F{}_{2,1} & K_F{}_{2,2} & ...   \\
			\vdotswithin{} & \vdotswithin{}& \vdotswithin{}& \vdotswithin{}& \vdotswithin{}&
		\end{pmatrix}\begin{pmatrix}
			1\\
			1\\
			1\\
			0\\
			\vdotswithin{}
		\end{pmatrix}\nonumber\\
		&=\begin{pmatrix}
			3\\
			3\\
			3+K_{F1,1}\\
			K_{F2,1}\\
			\vdotswithin{}
		\end{pmatrix}=2\begin{pmatrix}
			1\\
			1\\
			1\\
			0\\
			\vdotswithin{}
		\end{pmatrix}+\begin{pmatrix}
			1\\
			1\\
			1+K_{F1,1}\\
			K_{F2,1}\\
			\vdotswithin{}
		\end{pmatrix}\nonumber\\
		&= 2 \cdot l_{ele} +K_{B,3}.
	\end{align}
	Multiply $ K_B^{-1}$ on both sides,
	\begin{align}\label{id}
		K_B^{-1} l_{ele} =\frac{1}{2}(l_{ele}-\mathbf{e_3})=\frac{1}{2}\begin{pmatrix}
			1\\
			1\\
			0\\
			\vdotswithin{}
		\end{pmatrix}.
	\end{align}
	Therefore, the self statistics is 
	\begin{align}\label{eself}
		\theta_{ele} = \pi l_{ele}^T K_B^{-1} l_{ele} = \pi,
	\end{align} and the mutual statistics with any $l\in\mathbb{Z}^n$ is 
	\begin{align}\label{emut}
		\theta_{l,e} = 2\pi l^T K_B^{-1} l_{ele} = \pi (l_1+l_2).
	\end{align}
	Eq.\ref{eself} shows that, regardless of the exact form of $K_F$, $l_{ele}=(1,1,1,0,...)^T$ always behaves like a fermion. On the other hand, requiring Eq.\ref{emut} to be trivial, we found
	\begin{align}
		l_1+l_2 \in 2\mathbb{Z}
	\end{align}

	\subsection{Statistics is preserved under identification}\label{statpres}
	It suffices to prove that the lower right corner of $K_B$ is the inverse of $K_F. $ Indeed, the inverse of $K_B$ takes the explicit form,
	\begin{align}\label{kbinv}
	&K_B^{-1} = \nonumber\\
	&\begin{pmatrix}
	\frac{1}{4}(K_F^{-1})_{1,1}& \frac{1}{2}+\frac{1}{4}(K_F^{-1})_{1,1}& -\frac{1}{2}(K_F^{-1})_{1,1}&-\frac{1}{2}(K_F^{-1})_{1,2}&...\\
	\frac{1}{2}+\frac{1}{4}(K_F^{-1})_{1,1} & \frac{1}{4}(K_F^{-1})_{1,1}&  -\frac{1}{2}(K_F^{-1})_{1,1}&-\frac{1}{2}(K_F^{-1})_{1,2}&...\\
	-\frac{1}{2}(K_F^{-1})_{1,1}&	-\frac{1}{2}(K_F^{-1})_{1,1}& (K_F^{-1})_{1,1} & (K_F^{-1})_{1,2}&...\\
	-\frac{1}{2}(K_F^{-1})_{2,1}&	-\frac{1}{2}(K_F^{-1})_{2,1}&(K_F^{-1})_{2,1}&(K_F^{-1})_{2,2}&...\\
	\vdotswithin{}&\vdotswithin{}&\vdotswithin{}&\vdotswithin{}&\ddots
	\end{pmatrix}\\
&= \begin{pmatrix}
	\frac{1}{4}(K_F^{-1})_{1,1} & \frac{1}{2}+\frac{1}{4}(K_F^{-1})_{1,1} & -\frac{1}{2}(K_F^{-1})^T_1\\
	\frac{1}{2}+\frac{1}{4}(K_F^{-1})_{1,1} &\frac{1}{4}(K_F^{-1})_{1,1} &  -\frac{1}{2}(K_F^{-1})^T_1\\
	-\frac{1}{2}(K_F^{-1})_1 & -\frac{1}{2}(K_F^{-1})_1 & K_F^{-1}
\end{pmatrix},
	\end{align} where $(K_F^{-1})_1 $ is the first column of $K_F^{-1}$. One can then explicitly check that 
\begin{align}
	K_B^{-1}\cdot K_B =\mathbb{1}\nonumber.
\end{align} 

Notice that the lower right corner is exactly the inverse of $K_F$. Thus 
\begin{align}
&l_B^T K_B^{-1} l_B \nonumber\\
=& \begin{pmatrix}
	0 & 0 & l_f
\end{pmatrix}  \begin{pmatrix}
	\frac{1}{4}(K_F^{-1})_{1,1} & \frac{1}{2}+\frac{1}{4}(K_F^{-1})_{1,1} & -\frac{1}{2}(K_F^{-1})^T_1\\
	\frac{1}{2}+\frac{1}{4}(K_F^{-1})_{1,1} &\frac{1}{4}(K_F^{-1})_{1,1} &  -\frac{1}{2}(K_F^{-1})^T_1\\
	-\frac{1}{2}(K_F^{-1})_1 & -\frac{1}{2}(K_F^{-1})_1 & K_F^{-1}
\end{pmatrix} \begin{pmatrix}
0\\
0\\
l_f
\end{pmatrix} \nonumber\\
=& l_f^T K_F^{-1} l_f.
\end{align}

	\section{Gapped boundaries between $\mathbb{Z}_2$ and $\mathbb{Z}_2$}\label{bdryz2z2}
	
	Here we elaborate on the meaning of \emph{trivial boundary} between two $\mathbb{Z}_2$ topological orders. Consider the $K$ matrix
	\begin{align}
		K=\begin{pmatrix}
			0 & 2 & 0 & 0\\
			2 & 0 & 0 & 0\\
			0 & 0 & 0 & 2\\
			0 & 0 & 2 & 0\\
		\end{pmatrix} =\begin{pmatrix}
			K_{\mathbb{Z}_2} & 0\\
			0 & K_{\mathbb{Z}_2}
		\end{pmatrix}.
	\end{align} We found several gapped boundaries, such as
	\begin{align*}
		\{ \begin{pmatrix}
			0\\
			0\\
			0\\
			0
		\end{pmatrix}, \begin{pmatrix}
			0\\
			0\\
			1\\
			0
		\end{pmatrix}, \begin{pmatrix}
			1\\
			0\\
			0\\
			0
		\end{pmatrix}, \begin{pmatrix}
			1\\
			0\\
			1\\
			0
		\end{pmatrix}\}
	\end{align*} or 
	\begin{align*}
		\{ \begin{pmatrix}
			0\\
			0\\
			0\\
			0
		\end{pmatrix}, \begin{pmatrix}
			0\\
			0\\
			1\\
			0
		\end{pmatrix}, \begin{pmatrix}
			0\\
			1\\
			0\\
			0
		\end{pmatrix}, \begin{pmatrix}
			0\\
			1\\
			1\\
			0
		\end{pmatrix}\}.
	\end{align*} In particular, there exists a solution 
	\begin{align}
		S_{t.b.} &= \{ \begin{pmatrix}
			0\\
			0\\
			0\\
			0
		\end{pmatrix}, \begin{pmatrix}
			1\\
			1\\
			1\\
			1
		\end{pmatrix}, \begin{pmatrix}
			0\\
			1\\
			0\\
			1
		\end{pmatrix}, \begin{pmatrix}
			1\\
			0\\
			1\\
			0
		\end{pmatrix}\}\nonumber\\
		&=   \{ \begin{pmatrix}
			\mathbb{1}\\
			\mathbb{1}
		\end{pmatrix}, \begin{pmatrix}
			f\\
			f
		\end{pmatrix}, \begin{pmatrix}
			m\\
			m
		\end{pmatrix}, \begin{pmatrix}
			e\\
			e
		\end{pmatrix}\}.
	\end{align} Having the same excitation condensed on both sides of a boundary means that an excitations can pass through the boundary without changing their types. Thus, we suggest that $S_{t.b.}$ is the trivial gapped boundary.

\bibliography{main,../../../bib/all,../../../bib/publst,../../../bib/allnew,../../../bib/publstnew}

\end{document}